# Control of thin NbN film superconducting properties by ScN buffer layer

N.V. Porokhov[1]*, M.A. Dryazgov[1], A.R. Shevchenko[1], A.M. Mumlyakov[1], Z.S. Enbaev[1], Yu.V. Blinova[2], D.I. Devyaterikov[2], Yu.P. Korneeva[1], A.A. Korneev[1,3], M. A. Tarkhov[1].

[1]*Institute of Nanotechnology of Microelectronics of Russian Academy of Science, 119991, Russia*

[2]*M.N. Miheev Institute of metal physics UB RAS. Yekaterinburg, 620990, Russia*

[3]*Higher School of Economics - National Research University, Myasnitskaya 20, Moscow 101000, Russia*

*e-mail:* nporokhov@gmail.com

**Abstract**

This work investigates the effect of a scandium nitride (ScN) buffer layer on the superconducting properties of niobium nitride (NbN) thin films. The use of a ScN buffer layer significantly improves the characteristics of 29 nm-thick NbN films: the critical temperature $T_c$ increases from 9 K to 12.5 K, while the resistivity $\rho$ at 20 K decreases from 330 μΩ·cm to 210 μΩ·cm compared to films without a buffer layer. These enhancements are attributed to the better lattice matching between NbN and ScN, which results in a higher quality crystal lattice of the NbN film, as confirmed by transmission electron microscopy (TEM) and X-ray diffraction (XRD) data.

## 1. Introduction

Thin-film superconducting structures based on niobium nitride (NbN) are widely used in superconducting nanoelectronics devices, including single-photon detectors (SNSPD) [1,2], hot-electron bolometers (HEB) and terahertz mixers [3], as well as microwave kinetic inductance detectors (MKIDs) [4]. In many applications high resistivity Si substrates are favorable due to low dielectric losses as well as their compatibility with the CMOS technology. Unfortunately, lattice mismatch between Si and NbN decreases the superconducting transition temperature $T_c$ and increases resistivity $\rho$, especially in thin films. A fundamental approach to improving the superconducting properties of thin NbN involves epitaxial film growth. This can be achieved by using substrates or buffer layers with lattice parameters matching those of the deposited superconducting film.

In Refs. [5-7] a monocrystalline MgO substrate was used to improve the superconducting characteristics of NbN. The critical temperature was reported to be close to that of the bulk material [7] due to the proximity of the crystal lattice parameters: 0.421 nm for MgO and 0.439 nm for NbN. It is noteworthy that MgO was also widely used as a buffer layer on sapphire substrates enabling the integration of NbN with other materials in the realization of HEB devices [8-10] and Josephson tunnel junctions [11,12]. However, MgO is hygroscopic, which leads to problems in lithography and subsequent premature degradation of samples.

AlN, in turn, was demonstrated as a promising buffer layer for 8-nm-thick NbN [13] to increase the $T_C$ from 7.3 K to 10 K. High-quality polycrystalline NbN films were realized on 8-inch silicon wafers [14] with AlN buffer layer, and their feasibility for SNSPDs was demonstrated. However, due to the difference in thermal expansion coefficients [15], residual mechanical stresses are created in AlN, which negatively affect the width of the superconducting transition $\Delta T$, broadening it to 3 K for 8-nm-thick NbN film. The disadvantage of AlN as a buffer layer for NbN is the crystallographic mismatch between the crystal lattices: the hexagonal crystal structure of AlN versus the cubic lattice of NbN, which is a potential cause of dislocations at the buffer layer – superconducting film interface. In [16] 10-nm-thick AlN buffer layer increased $T_C$ from 5.0 K to 7.2 K and from 7.3 K to 9.3 K for NbN layers with thicknesses of 5 nm and 7 nm, respectively. In this work, the X-ray diffraction method (XRD) and transmission electron microscopy (TEM) explained the improvement in superconducting characteristics by a change in the crystalline structure of NbN.

Another buffer layer that provides an increase in $T_C$ by 1-3 K for NbN films compared to those grown on silicon substrates is TiN [17], due to a reduction in the thickness of the defective interfacial NbN layer, as well as an improvement in the crystalline orientation and grain size of NbN. TEM studies presented in the work showed that NbN "inherits" the crystalline structure of TiN, which on one hand increases $T_C$ by increasing the grain size to 75 nm (NbN/TiN/Si) compared to 20 nm (NbN/Si). Furthermore, atomic force microscopy (AFM) results presented in the work showed that the surface roughness of NbN/TiN/Si is higher than that of the NbN/Si film, namely 0.8 nm versus 0.5 nm. The disadvantages of TiN include its ability to absorb oxygen, which can potentially diffuse into NbN, requiring more careful control of the deposition process parameters, particularly the $N_2$/Ar flow ratio. Nevertheless, epitaxial three-layer NbN/AlN/NbN structures on silicon substrates with a TiN buffer layer demonstrated excellent tunneling properties [18].

In [19], GaN/AlN was used for growth of 8-nm-thick NbN film with $T_C$ = 13.1 K. Nevertheless, GaN is a challenging material for use as a buffer layer for NbN due to its significantly smaller lattice parameter compared to NbN (GaN a ≈ 3.189 Å) and its hexagonal crystal structure. Later studies [20] investigated the influence of annealing as a method for improving the crystallinity of the NbN film and increasing the grain size. In this study GaN did not prove effective: the $T_C$ was 8.9 K, while on other nitrides (SiN, AlN) it was above 10.2 K. After annealing at 900°C, the NbN on the GaN buffer layer almost completely peeled off from the substrate, possibly due to relatively low adhesion on the GaN buffer layer compared to other substrates, which could also lead to poor superconductivity. In the same work, it was also shown that a SiN buffer layer, like AlN, improves the quality of NbN film crystallization but results in a smaller grain size compared to AlN and a much greater surface roughness: 29 nm for SiN versus 5 nm for AlN. AlN, in turn, demonstrated the best result after annealing at 900°C: the lattice constant of NbN approached 0.439.

In [21], VN on sapphire was studied as a buffer layer for NbN. The researchers proposed an optimal deposition temperature of 400°C, at which a "step flow" film growth mode is demonstrated, leading to an atomically smooth surface with $T_C$ = 16.6 K, close to the bulk material. However, when the temperature is increased to 800°C, the achievable $T_C$ drops to approximately 10 K due to the appearance of the non-superconducting hexagonal phase $\delta^o$-NbN. According to AFM studies in the same work, small columnar structures (below 10 nm) begin to grow on the steps of NbN, which increases roughness and can serve as a source of defects.

SiC as a buffer layer possesses very high crystalline quality and thermal stability. In [22], 3.4 – 4.1-nm-thick NbN films were deposited at a temperature 800°C on 3C-SiC/Si and $T_C$ = 11.8 K was achieved. Later, in [23], the operation of NbN terahertz HEB mixers on SiC buffer layer was demonstrated.

The ScN buffer layer, proposed in this work, offers the following advantages: i) it is chemically and thermally stable and non-hygroscopic; ii) it has the same cubic crystal structure (NaCl type) as NbN; iii) it is a dielectric, providing ideal electrical insulation between the substrate and NbN; iv) it has a lattice constant very close to that of NbN: 0.450 nm for ScN versus 0.439 nm for NbN. Furthermore, the films were fabricated using a room-temperature deposition process. All improved superconducting properties were achieved in a 30 nm thick NbN film deposited on ScN at room temperature.

This work investigates the influence of a ScN buffer layer (up to 38 nm thick) on the superconducting properties of 29 nm-thick NbN films. This thickness of NbN was chosen from the following considerations: ScN is a new buffer layer for NbN with its influence on NbN being unknown. Thus, increased thickness ensures reasonable superconducting properties of NbN. Another consideration is the limitation of XRD method which we use for microstructural analysis: for proper accuracy it requires a rather thick film. The results demonstrate that increasing the ScN thickness improves the superconducting characteristics of NbN: the critical temperature rises from 9 K to 12.5 K, while the resistivity at 20 K decreases from 330 µΩ·cm to 210 µΩ·cm. Microstructural analysis of NbN films reveals that when deposited on ScN layers thicker than 15-20 nm, the NbN films inherit the crystalline structure and grain orientation of the ScN buffer layer. Moreover, the lattice parameter of NbN approaches that of ScN.

## 2. Fabrication process

In this study, we fabricated a series of experimental samples with a multilayer n.Si/SiO$_2$/ScN/NbN structure (Figure 1(a)). The substrates were of 4-inch low-resistivity n-type monocrystalline silicon wafers with a pre-grown 500 nm thermal oxide layer. A ScN buffer layer was deposited onto this amorphous SiO$_2$ surface by reactive magnetron sputtering, followed by in situ NbN layer deposition. The ScN thickness varied for different samples from 3.6 to 38 nm, while the NbN thickness remained constant at 29 nm. High-purity (99.95%) metallic scandium and niobium targets were used for deposition. The process was conducted at room temperature and a process chamber pressure of 0.2 Pa in a mixture of argon and nitrogen (4:1 ratio) with a DC magnetron power density of 5 W/cm$^2$, yielding a deposition rate of 11 nm/min. For comparison, reference NbN films were deposited on bare SiO$_2$ substrates under identical conditions.

It has been reported previously that ScN films deposited by magnetron sputtering show semiconducting behavior, and their resistivity increases with decreasing thickness. For thicknesses of 480-520 nm, the resistivity is $\rho = 0.22$ mΩ·cm [24]; at 300 nm, $\rho = 2$ mΩ·cm [25]; and for thicknesses below 180 nm, $\rho = 11 - 16\ m$Ω·cm [26]. A recent study [27] reported investigations of 10-40 nm thick ScN films with resistivities up to $17.9$ mΩ·cm. In our experiments, the films either exhibited dielectric behavior at room temperature or showed increasing resistance upon cooling, eventually becoming dielectric at helium temperatures.

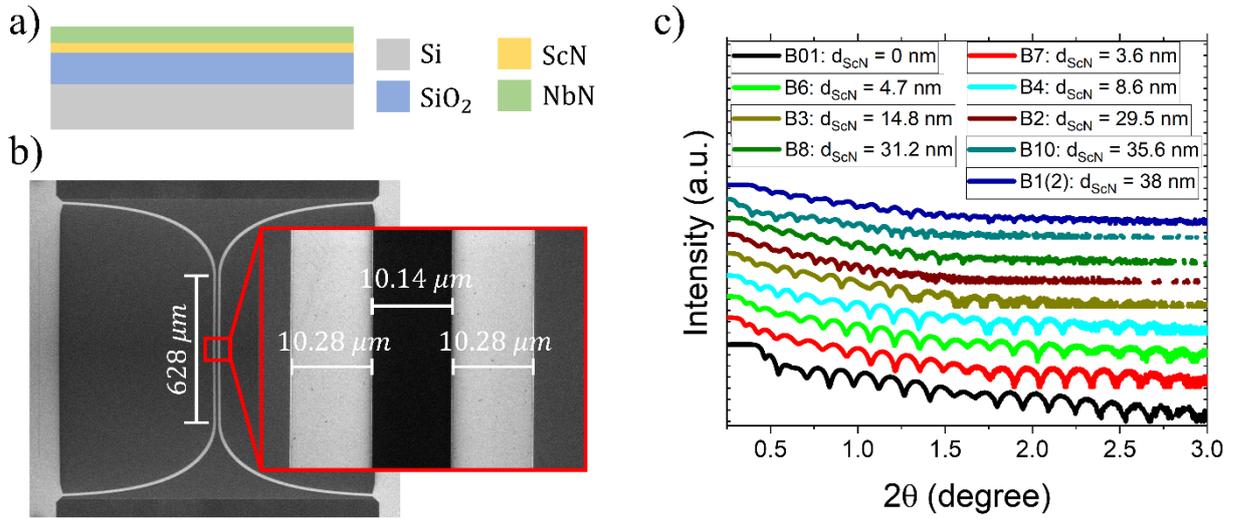

**Figure 1 (a)** Schematic layer sequence. **(b)** Micrograph of samples for transport measurements; inset shows an enlarged central region. **(c)** Experimental X-ray reflectometry (XRR) results for sample series B0–B10.

The microstructural properties were studied on unpatterned films. The film thickness was determined using X-ray reflectometry (XRR). To investigate the transport properties, microbridge structures with a width of $w_{strip} = 10$ μm and length of 628 μm were fabricated (see SEM image in Figure 1(b)) by optical lithography and inductively coupled plasma reactive ion etching (ICP-RIE). The process was carried out at a pressure of 10 mTorr with a SF$_6$/Ar (2:1) gas chemistry. The plasma was generated with an ICP source power of 430 W and a RF bias power of 15 W was applied to the substrate holder, which was maintained at 50 °C. The etch time was 1 minute.

The layer thicknesses and densities were determined using X-ray reflectometry (XRR). Measurements were performed on an Empyrean diffractometer (Malvern Panalytical) with monochromatic Cu-Kα radiation (λ = 1.5406 Å). Prior to analysis, the samples underwent ultrasonic cleaning in isopropanol and surface leveling to minimize measurement artifacts. Figure 1(c) shows the X-ray reflectometry patterns of the sample series with varying $d_{ScN} = 0 - 38$ nm. The obtained thickness values are presented in Table 1.

Analysis of X-ray reflectometry data revealed a surface niobium oxide layer (Nb$_x$O$_y$) with reduced density in all samples. Its presence was established by fitting the XRR data using a multilayer model. This model included a top layer with a reduced electron density (~4.5 g/cm³), which is consistent with NbO$_x$. Although bulk NbN is considered hard for oxidation, Nb$_x$O$_y$ on the surface of thin-film NbN is consistently observed in many studies, see e.g. [28]. This oxide layer forms naturally through interaction of the thin film surface with ambient oxygen. The consistent NbN layer thickness (29 nm) across all specimens demonstrates the high reproducibility of the magnetron sputtering process. Measured parameters of the samples are presented in Table 1.

Table 1. Sample parameters (sorted by ScN thickness, $d_{ScN}$)

| Sample ID | $d_{ScN}$ (nm) | $R_s(20K)$ ($\Omega/\square$) | $\rho^{20K}$ ($\mu\Omega\cdot cm$) | RRR | $T_c$ (K) | D ($cm^2/s$) | $I_c(2.5K)$ (mA) | $a_{NbN}$ (Å) | $a_{ScN}$ (Å) |
|---|---|---|---|---|---|---|---|---|---|
| B01 | 0 | 113.9 | 330.3 | 0.77 | 9.2 | | 8.79 | 4.339 | |
| B7 | 3.6 | 91.5 | 265.4 | 0.74 | 9.9 | 0.43 | 11.5 | 4.372 | 4.447 |
| B6 | 4.7 | 93.2 | 270.3 | 0.72 | 10.2 | | 12.5 | 4.372 | 4.450 |
| B4 | 8.6 | 82.7 | 239.8 | 0.74 | 11 | 0.46 | 12 | 4.389 | 4.447 |
| B3 | 14.8 | 73.9 | 214.3 | 0.77 | 12.5 | | 18 | 4.384 | 4.450 |
| B2 | 29.5 | 74.6 | 216.3 | 0.76 | 12.6 | | 23 | 4.381 | 4.505 |
| B8 | 31.2 | 73.3 | 212.6 | 0.79 | 12.8 | | 25.6 | | |
| B1 | 34 | 74.3 | 215.5 | 0.78 | 12.4 | | 22.6 | 4.378 | 4.505 |
| B10 | 35.6 | 71.1 | 206.2 | 0.84 | 12.9 | | 21.9 | | |
| B1(2) | 38 | 73.7 | 213.7 | 0.79 | 12.2 | 0.47 | 22.6 | | |

## 3. Transport measurements

Low-temperature transport measurements were performed in a Gifford-McMahon cryocooler with the base temperature of 2.6 K. DC measurements employed a Keithley 2460 SourceMeter as the current/voltage source. Temperature-dependent measurements (in 2.6–20 K range) utilized a heater mounted on the cryocooler's cold plate, controlled by a LakeShore 325 temperature controller.

The dependence of resistance on temperature $R(T)$ was measured on the patterned samples in constant-current mode at 10 μA probe current. The critical temperature ($T_c$) was determined as the temperature corresponding to the maximum value of the derivative $dR/dT$. The resistivity at 20 K ($\rho^{20K}$) and sheet resistance ($R_s$) were calculated from the microbridge resistance measured at 20 K and its geometric dimensions. Critical currents were extracted from current-voltage characteristics measured in voltage-source mode, with $I_c$ defined as the maximum current at the switching point where a normal domain nucleates, causing abrupt current collapse under fixed source voltage conditions.

Figure 2(a) shows the dependences of $T_c$ and $\rho^{20K}$ on the ScN thickness ($d_{ScN}$). The leftmost data points at $d_{ScN} = 0$ correspond to the reference samples without ScN. The addition of the ScN layer leads to an increase in critical temperature and a decrease in resistivity compared to the reference sample values. It should be noted that the residual resistance ratio $RRR = R^{300K}/R^{20K}$ is independent of ScN thickness, as shown in Figure 2(b). The RRR value remains approximately 0.8, indicating an unchanged internal structure of NbN.

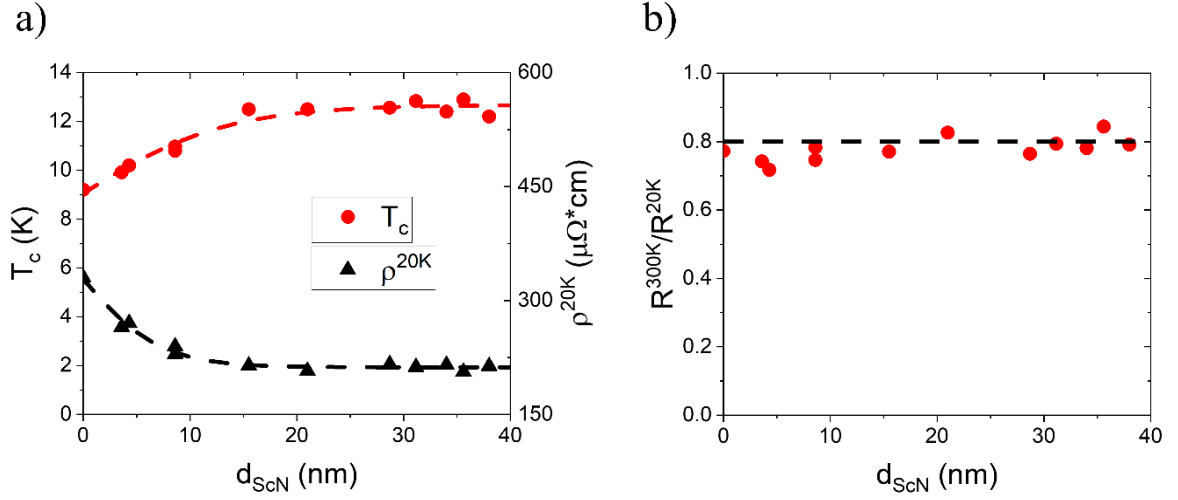

**Figure 2.** (a) Measured dependences of the critical temperature ($T_c$) and the resistivity at 20 K ($\rho^{20K}$) on the thickness of ScN layer ($d_{ScN}$), the dashed lines are the guides for the eye. (b) The dependence of the residual resistance ratio on the thickness of ScN layer ($d_{ScN}$), the dashed line corresponds to $RRR \approx 0.8$.

Figure 3 presents the temperature dependences of the critical current $I_C(T)$ for all measured samples, normalized to their $T_c$ and $I_c(0K)$ values. The dashed line represents a fit using the two-fluid model expression [28]:

$$I_C(T) = I_C(0)(1 - (T/T_C)^2)^{3/2}(1 + (T/T_C)^2)^{1/2}. \qquad (1)$$

This dependence describes the $I_c(T)$ behavior near $T_c$ by the $(1 - T/T_C)^{3/2}$ according to Ginzburg-Landau theory and saturates as $T \to 0$. Here, $I_c(0)$ serves as fitting parameter, and $T_c$ is taken from $R(T)$ mesurements. In literature, there are different $I_c(T)$ dependencies observed for different types of samples (see e.g. [29] and [30]) caused by different physical mechanisms such as vortex penetration and motion or junction-like inter-grain boundaries. In our case, the data show that the mechanism of the critical current is the same for different thicknesses of ScN. This dependence also allowed us to extrapolate the $I_c$ values to 0 K temperature and compare them with the calculated values of the depairing currents.

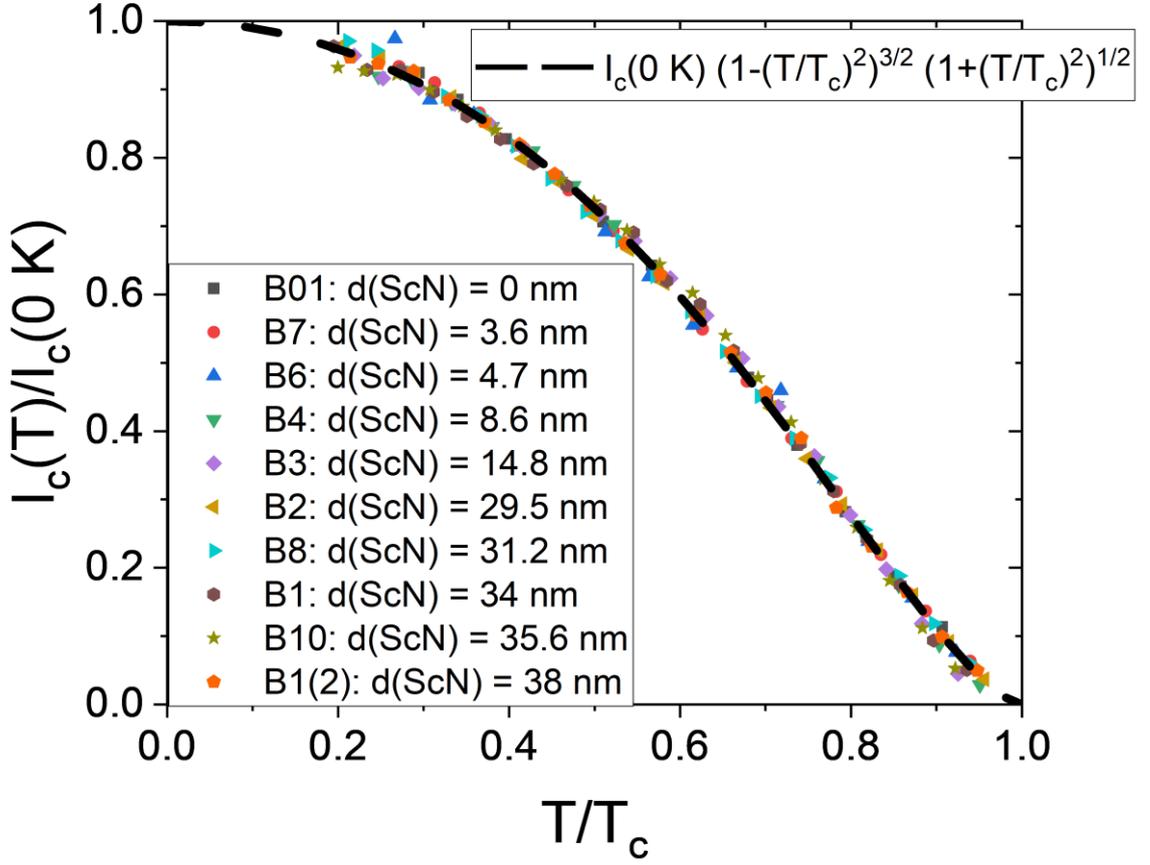

**Figure 3** Experimentally obtained temperature dependences of the critical current, normalized to $T_c$ and $I_c(0K)$ values. The dashed line indicates the theoretically predicted trend according to Equation (1).

Figure 4(a) shows the dependences of $I_c(0\text{ K})$ on ScN layer thickness, along with the predicted depairing current values at $T = 0\text{ K}$ calculated with Eq 21 from [31], where we substitute $N(0) = (2e^2 D R_s d)^{-1}$:

$$I_{dep}(0) = 0.74 \frac{w_{strip}(\Delta(0))^{\frac{3}{2}}}{e R_S \sqrt{\hbar D}}. \qquad (2)$$

The values of $R_s$ and $T_C$ in Equation 2 are experimental data for each sample. The diffusion coefficient $D$ was experimentally determined for several samples (B7, B4, and B1(2)) from temperature dependence of the critical magnetic field $H_{C2}$ with the common formula:

$$D = \frac{4k_B}{\pi e}\left[\frac{dH_{C2}}{dT}\right]^{-1} \qquad (3)$$

(Some details how this formula is derived can be found e.g. in [32]). In Figure 4(a) these samples are indicated by green. For the remaining samples, the average value of $D = 0.454 \text{ cm}^2/\text{s}$ was used. Notably, $I_c(0 \text{ K})/I_{dep}(0 \text{ K}) \approx 0.47$ for all samples, as shown in Figure 4(b). The independence of the $I_c/I_{dep}$ ratio from ScN layer thickness indicates preserved structural homogeneity of the NbN film. Furthermore, the proximity of the critical current to the depairing current $I_C \approx 0.5 I_{dep}$. For 5-nm-thick NbN (usually used for SNSPD) such a ratio is favorable for single-photon detection in wide range of photon energies (see inset of Fig. 11 in [33], where $\gamma = 10$ corresponds to NbN). If we assume as a rough estimation that in the thick film a photon produces the same number of quasiparticles, then we may expect that hotspot radius scales as $1/\sqrt{d}$. Thus, for 29-nm-thick film its radius is ~2.2 times smaller compared to 5-nm-thick film. And since SNSPD usually does not demonstrate sharp cut-off wavelength, we may hope that even such thick film are capable of detecting high-energy photons.

Critical current reaches its maximum values when ScN thickness $d_{ScN}$ is higher than 15-20 nm. The maximum critical current density calculated considering the microbridge cross-section is about 8 MA/cm².

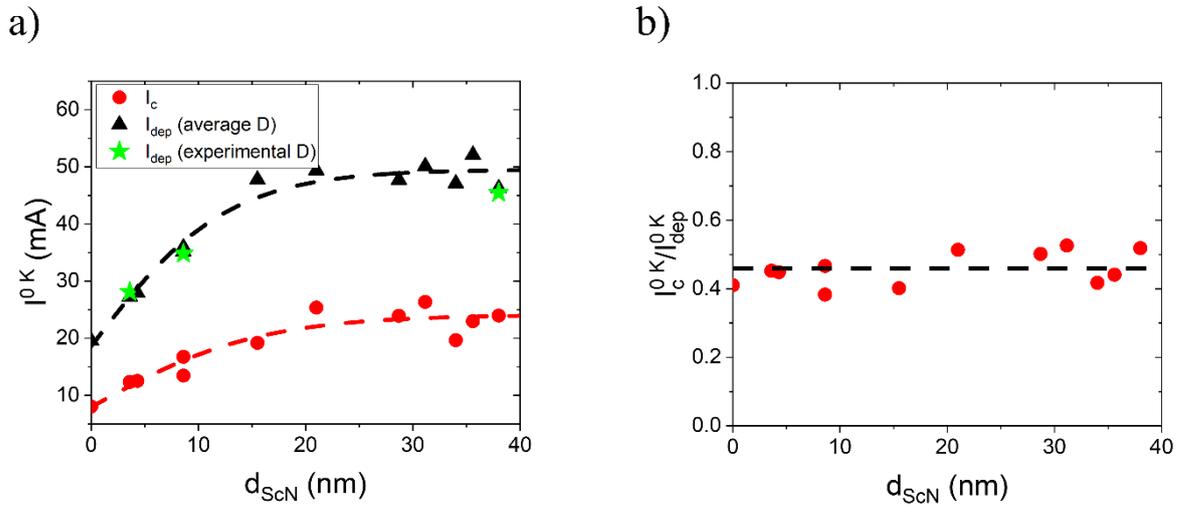

**Figure 4** (a) Experimentally obtained critical current values $I_c$ (red circles) and theoretically calculated depairing current values $I_{dep}$ (black triangles) at 0 K for samples with different ScN thicknesses $d_{ScN}$, the dashed lines are guides to the eye. (b) The ratio of measured critical current to calculated depairing current at 0 K, the dashed line corresponds to $I_c/I_{dep} \approx 0.47$.

### 4. Discussion and microstructural analysis

The film microstructure was examined by transmission electron microscopy (TEM) using a JEOL JEM-2100 Plus microscope operating at 200 kV. Cross-sectional lamellae for TEM analysis were prepared using a focused ion beam (FIB). Figure 5 shows TEM images of samples without ScN buffer layer and with it. The layer sequences are as follows: (a) bottom silicon oxide layer, dark high-density ~29 nm NbN layer, and light-contrast aluminum layer; (b) bottom silicon oxide layer, light 3.6 nm ScN buffer layer, dark high-density NbN layer, surface layer (<3 nm), and top aluminum contrast layer; (c) bottom silicon oxide layer, light 30 nm ScN layer, dark high-density NbN layer, surface layer (<3 nm), and top aluminum contrast layer; (d) bottom ScN layer and dark high-density NbN layer. Images (b), (c), and (d) in Figure 5 reveal grain formation penetrating through both ScN and NbN layers, resembling a columnar structure.

The 30 nm-thick ScN layer in Figure 5(c) exhibits a well-defined columnar structure with an average column width of ~5 nm. This columnar morphology, observed only for ScN buffer layer thicknesses ≥ 10 nm, is characteristic of films grown by magnetron sputtering at low homologous temperatures ($T/T_m$), consistent with Thornton's Structure Zone Model (Zone $1/T$) [34]. The formation and development of this nanocolumnar structure occurs beyond a critical thickness where competitive growth and geometric shadowing mechanisms become dominant. The columnar structure of ScN is effectively transferred to the epitaxially growing NbN layer, creating a continuous nanocolumnar architecture throughout the entire heterostructure. Importantly, the dense nature of the intercolumnar boundaries (without any pronounced voids visible) in this nanocrystalline structure preserves the superconducting properties, as confirmed by the high critical temperature of our NbN films. This ScN columnar structure was previously confirmed by TEM studies in reference [26], though for films with 10-fold greater thickness ($d_{ScN}$ = 345 nm).

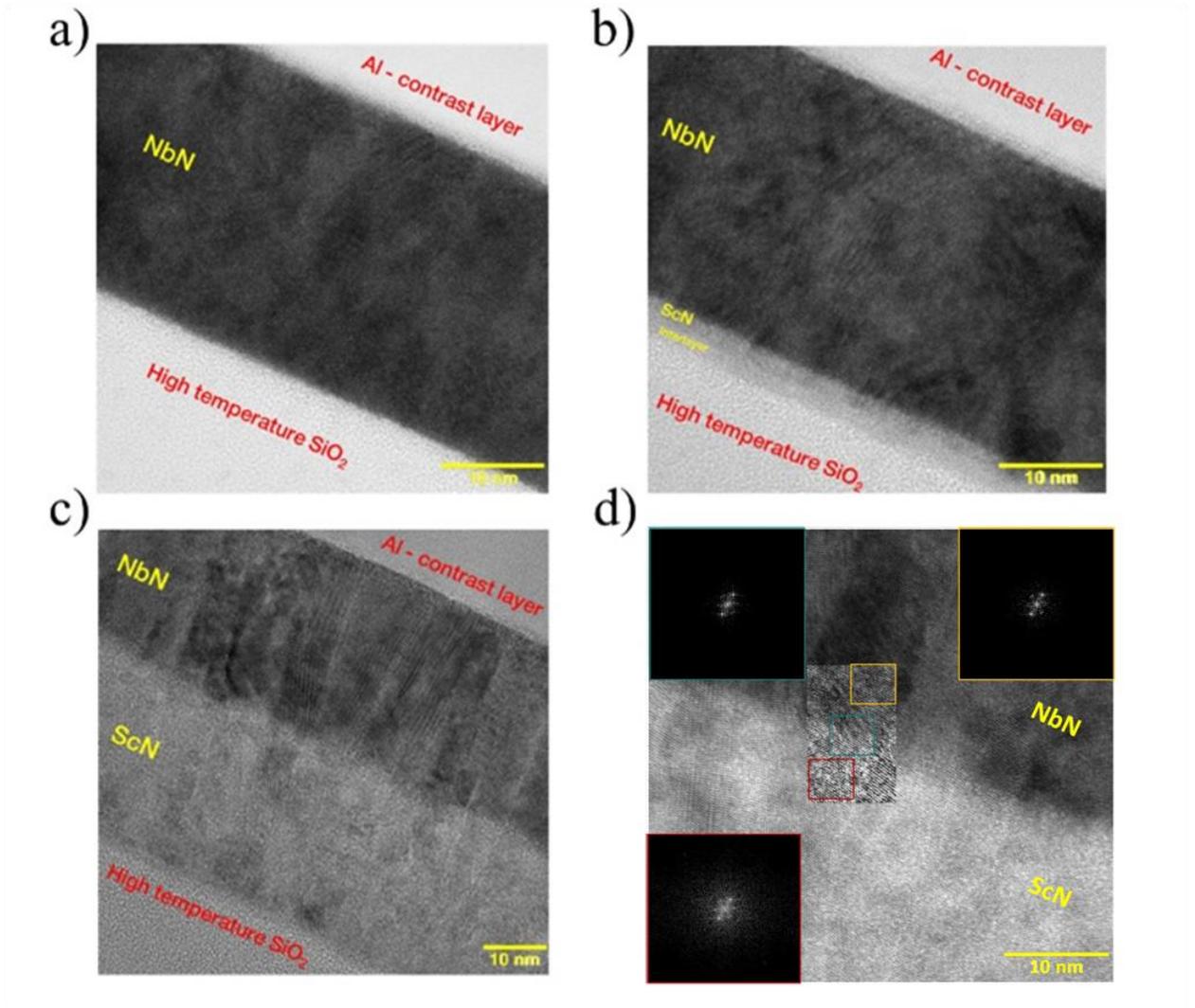

**Figure** 5 Cross-sectional TEM images of: (a) sample without ScN; (b) sample with $d_{\text{ScN}} = 3.6$ nm buffer layer; (c) sample with $d_{\text{ScN}} = 30$ nm buffer layer; (d) ScN/NbN interface region.

Notably, the maximum $T_c > 12$ K is achieved when $d_{\text{ScN}}$ is larger than 15 nm. This enhancement may result from modified NbN crystal structure, where the buffer layer improves crystal lattice of NbN and induces long-range order by replicating the ScN layer's crystal structure and orientation within columnar crystallites. Similar behavior was demonstrated for thin (≤10 nm) NbN films deposited on AlN layer [16]. The absence of an interfacial layer at the ScN/NbN boundary (Figure 5(d)) directly results from the in-situ deposition process that prevents atmospheric exposure between growth stages.

Figure 6 presents X-ray diffractograms obtained by grazing incidence X-ray diffraction (GIXRD) using Co Kα radiation (λ $_{\text{Kα1}}$=1.7890 Å, λ $_{\text{Kα2}}$=1,7929 Å) on an Empyrean Panalytical diffractometer, with an ω angle of 1° between the primary beam and sample surface. The observed diffraction peaks can be indexed as 111, 200, 220, 311, and 222 reflections of the NbN phase (space group 225). The 111 and 200 peaks of ScN phase overlap with corresponding NbN peaks and remain unresolved. The 220 peaks of ScN and NbN phases are resolved for samples B1 and B2. Lattice parameters of NbN and ScN determined by full-profile analysis are summarized in Table 1.

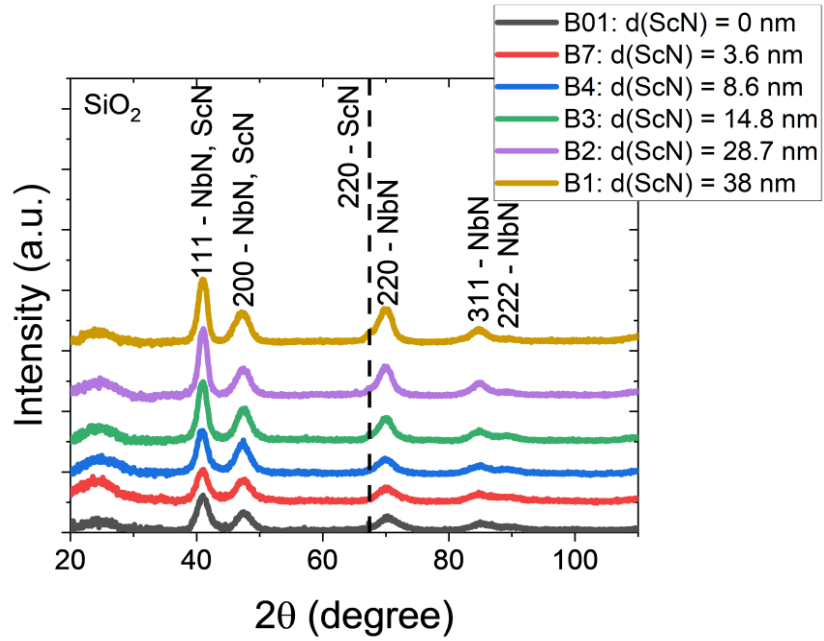

**Figure 6** X-ray diffractograms acquired by grazing-incidence XRD (GIXRD). The labeled peaks correspond to the indicated phases.

The inset of Figure 7(a) shows the dependence of the lattice parameter $a_0$ for NbN on the ScN layer thickness. The addition of ScN leads to an increase in $a_0$ from 4.34 Å (NbN without ScN, leftmost point at $d_{ScN} = 0$) to nearly 4.39 Å at $d_{ScN} = 8.5$ nm, followed by a slight decrease. Figure 7(b) presents the dependences of $T_C$ and $\rho^{20K}$ on $a_0$ for NbN. Qualitatively, they reproduce the known dependences on variations in the NbN lattice parameter obtained by changing the deposition parameters of NbN films [5], with maximum $T_C$ and minimum $\rho^{20K}$ achieved at $a_0 \approx 4.38$Å.

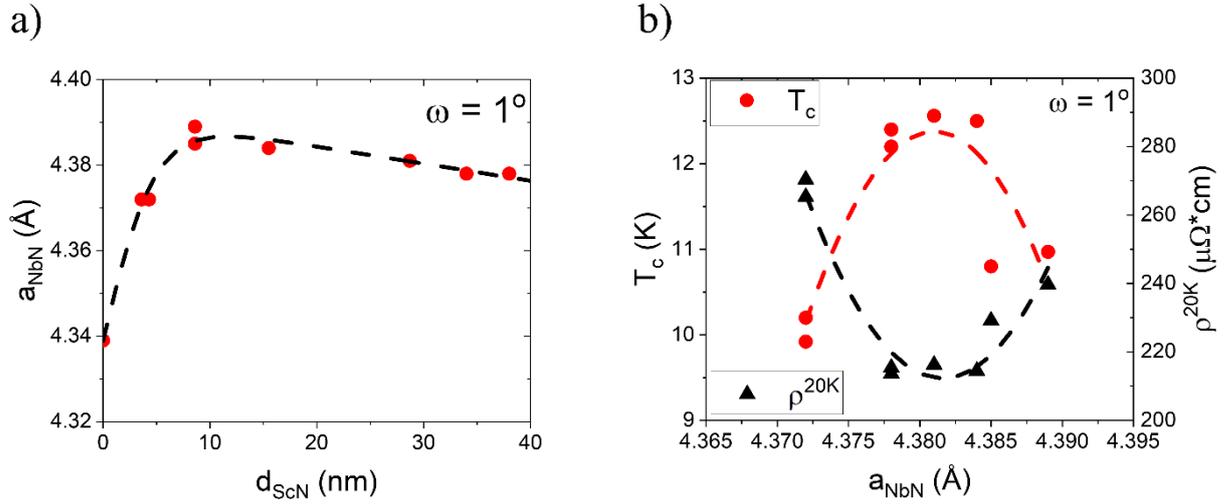

**Figure 7** (a) Dependences of the critical temperature ($T_C$) and resistivity at 20 K ($\rho^{20K}$) on the lattice parameter ($a_0$) for NbN. (b) The dependence of $a_0$ on ScN layer thickness ($d_{ScN}$). In all plots the dashed lines are the guides for the eye.

## 5. Conclusions

In this work, we have demonstrated that ScN buffer layer deposited by DC magnetron sputtering significantly enhances the superconducting properties of NbN films grown in-situ on amorphous SiO$_2$/Si substrates. The use of the ScN interlayer ensures a substantial improvement in the $T_C$ and $I_C$ of the NbN films. In the maximum enhancement $T_C$ reaches 12.5 K and at the minimum resistivity of 210 μΩ·cm, which are achieved for NbN film deposited on a 20-nm-thick ScN layer. This improvement in superconducting properties is directly correlated with the enhanced crystalline quality of NbN. XRD and TEM analyses confirm that the ScN layer promotes the growth of a columnar polycrystalline structure in NbN with a preferred orientation, in contrast to the randomly oriented grains in films deposited directly on silicon. The underlying mechanism is attributed to the close lattice matching between ScN and NbN, which facilitates epitaxial growth. This effect of underlying buffer layer crystal structure replication becomes pronounced when the ScN layer thickness is higher 15 nm. We believe our results establish ScN as an effective buffer layer for the integration of high-performance NbN superconductors with mainstream silicon technology.

### Acknowledgements

The study was supported by project No. 125020501540-9 of the Ministry of Education and Science of the Russian Federation. Fabrication and technology characterization were carried out at large scale facility complex for heterogeneous integration technologies and silicon+carbon nanotechnologies. The authors acknowledge L. Volkova for TEM sample preparation, M. Polyakova for TEM microstructure analysis, and D. Vodolazov for deep and useful consultations. The GIXRD study was carried out within the framework of the state assignment of the Ministry of Science and Higher Education of the Russian Federation for the IMP UB RAS.**References**

[1] Lau JA, Verma VB, Schwarzer D, Wodtke AM. 2023 Superconducting single-photon detectors in the mid-infrared for physical chemistry and spectroscopy. Chem Soc Rev. 52(3):921-941. DOI: 10.1039/d1cs00434d. PMID: 36649126.

[2] Natarajan C.M., Tanner M.G., and Hadfield R.H. 2012 Superconducting nanowire single-photon detectors: physics and applications Superconductor Science and Technology 25 063001. DOI:10.1088/0953-2048/25/6/063001,

[3] Klapwijk T.M. and Semenov A.V. 2017 Engineering physics of superconducting hot-electron bolometer mixers IEEE Transactions on Terahertz Science and Technology. 7(6) p 627. DOI: 10.1109/TTHZ.2017.2758267

[4] Seiichiro Ariyoshi, Kensuke Nakajima, Atsushi Saito, Tohru Taino, Hiroyuki Tanoue, Kensuke Koga, Noboru Furukawa, Hironobu Yamada, Shigetoshi Ohshima, Chiko Otani, and Jongsuck Bae1 2013 NbN-Based microwave kinetic inductance detector with a rewound spiral resonator for broadband terahertz detection Applied Physics Express Volume 6, Number 6. doi:10.7567/APEX.6.064103

[5] Wang, Z., Kawakami, A., Uzawa, Y., and Komiyama, B., 1996 Superconducting properties and crystal structures of single-crystal niobium nitride thin films deposited at ambient substrate temperature Journal of Applied Physics, vol. 79, no. 10, AIP, pp. 7837–7842, 1996. doi:10.1063/1.362392

[6] K. Zhang, K. Balasubramanian, B.D. Ozsdolay, C.P. Mulligan, S.V. Khare, W.T. Zheng, D. Gall 2016 Growth and mechanical properties of epitaxial NbN(001) films on MgO(001), Surface and Coatings Technology, Volume 288, Pages 105-114, https://doi.org/10.1016/j.surfcoat.2016.01.009

[7] Chockalingam, S.P.; Chand, M.; Jesudasan, J.; Tripathi, V.; Raychaudhuri, P. 2008 Superconducting properties and Hall effect of epitaxial NbN thin films. Phys. Rev. B 77 214503. DOI: 10.1103/PhysRevB.77.214503

[8] A. Kawakami, Zhen Wang and S. Miki, 2001 Low-loss epitaxial NbN/MgO/NbN trilayers for THz applications," in IEEE Transactions on Applied Superconductivity, vol. 11, no. 1, pp. 80-83, DOI: 10.1109/77.919289

[9] S. Miki, M. Fujiwara, M. Sasaki and Z. Wang, 2007 NbN Superconducting Single-Photon Detectors Prepared on Single-Crystal MgO Substrates," in IEEE Transactions on Applied Superconductivity, vol. 17, no. 2, pp. 285-288, DOI: 10.1109/TASC.2007.898582

[10] B. Guillet, V. Drakinskiy, R. Gunnarsson, O. Arthursson, L. Méchin, S. Cherednichenko, Y. Delorme, J.M. Krieg 2007 Influence of substrates and buffer layers on the quality of NbN ultra thin film for THz HEB, in Eighteenth International Symposium on Space Terahertz Technology, p. 153.

[11] Masahiro Aoyagi et al 1992 NbN/MgO/NbN Josephson Junctions for Integrated Circuits Jpn. J. Appl. Phys. 31 1778 DOI 10.1143/JJAP.31.1778

[12] Wang,H.Terai, W.Qiu,K.Makise,Y. Uzawa,K.Kimoto,Y. Nakamura 2013 High-quality epitaxial NbN/AlN/NbN tunnel junctions with a wide range of current density, Appl. Phys. Lett. 102, 142604 https://doi.org/10.1063/1.4801972

[13] Tatsuya Shiino, Shoichi Shiba, Nami Sakai, Tetsuya Yamakura, Ling Jiang, Yoshinori Uzawa, Hiroyuki Maezawa and Satoshi Yamamoto 2010 Improvement of the critical temperature of superconducting NbTiN and NbN thin films using the AlN buffer layer Supercond. Sci. Technol. 23 045004 DOI: 10.1088/0953-2048/23/4/045004

[14] Rhazi R. Machhadani H. Bougerol C. Lequien S. Robin E. Rodriguez G. Souil R. Thomassin J.-L. Mollard N. Désières Y. et al. 2021 Improvement of critical temperature of niobium nitride deposited on 8-inch silicon wafers thanks to an aln buffer layer. Supercond. Sci. Technol. 34, 045002. DOI: 10.1088/1361-6668/abe35e

[15] Hu, Z.; Pei, Y.; Fan, Q.; Ni, X.; Gu, X. 2025 Research on the Correlation of Physical Properties Between NbN Superconducting Thin Films and Substrates. Coatings, 15, 513. https://doi.org/10.3390/coatings15050513

[16] Valentin Brisson, Edith Bellet-Amalric, Nicolas Bernier, Jonathan Faugier-Tovar, Matthew Bryan, Joël Bleuse, Jean-Michel Gérard* and Ségolène Olivier 2025 Enhancement of the superconducting critical temperature of NbN ultra-thin films in a CMOS-compatible silicon nitride photonic platform on 200 mm-diameter wafers Supercond. Sci. Technol. 38 045018 DOI 10.1088/1361-6668/adbfc9

[17] J J Zhang, Xiaodong Su, L Zhang, L Zheng, X F Wang and Lixing You 2013 Improvement of the superconducting properties of NbN thin film on single-crystal silicon substrate by using a TiN buffer layer. Supercond. Sci. Technol. 26 045010 DOI 10.1088/0953-2048/26/4/045010

[18] Rui Sun; Kazumasa Makise; Lu Zhang; Hirotaka Terai; Zhen Wang 2016 Epitaxial NbN/AlN/NbN tunnel junctions on Si substrates with TiN buffer layers AIP Advances 6, 065119 https://doi.org/10.1063/1.4954743

[19] Diane Sam-Giao, Stéphanie Pouget, Catherine Bougerol, Eva Monroy, Alexander Grimm, Salha Jebari, Max Hofheinz, J.-M. Gérard, Val Zwiller; 2014 High-quality NbN nanofilms on a GaN/AlN heterostructure. AIP Advances; 4 (10): 107123. https://doi.org/10.1063/1.4898327

[20] Pei, Y.; Fan, Q.; Ni, X.; Gu, X. 2024 Controlling the Superconducting Critical Temperature and Resistance of NbN Films through Thin Film Deposition and Annealing. Coatings, 14, 496. https://doi.org/10.3390/coatings14040496

[21] John H. Goldsmith; Ricky Gibson; Tim Cooper; Thaddeus J. Asel; Shin Mou; Dave C. Look; John S. Derov; Joshua R. Hendrickson 2018 Influence of nitride buffer layers on superconducting properties of niobium nitride, J. Vac. Sci. Technol. A 36, 061502 https://doi.org/10.1116/1.5044276

[22] J. R. Gao, M. Hajenius, F. D. Tichelaar, T. M. Klapwijk, B. Voronov, E. Grishin, G. Gol'tsman, C. A. Zorman, M. Mehregany; 2007 Monocrystalline NbN nanofilms on a 3C-SiC/Si substrate. Appl. Phys. Lett.; 91 (6): 062504. https://doi.org/10.1063/1.2766963

[23] D Dochev, V Desmaris, A Pavolotsky, D Meledin, Z Lai, A Henry, E Janzén, E Pippel, J Woltersdorf and V Belitsky 2011 Growth and characterization of epitaxial ultra-thin NbN films on 3C-SiC/Si substrate for terahertz applications, Supercond. Sci. Technol. 24 035016 DOI 10.1088/0953-2048/24/3/035016

[24] P. V. Burmistrova, J. Maassen, T. Favaloro, B. Saha, S. Sala mat, Y. R. Koh, M. S. Lundstrom, A. Shakouri, and T. D. Sands, 2013 Thermoelectric properties of epitaxial ScN films deposited by reactive magnetron sputtering onto MgO(001) substrates J. Appl. Phys. 113 153704 DOI: 10.1063/1.4801886

[25] Bidesh Biswas, Bivas Saha 2019 Development of semiconducting ScN Phys. Rev. Materials 3, 020301 DOI:10.1103/PhysRevMaterials.3.020301

[26] D. Gall and I. Petrov 1998 Microstructure and electronic properties of the refractory semiconductor ScN grown on MgO(001) by ultra-high-vacuum reactive magnetron sputter deposition J. Vac. Sci. Technol. A 16 2411. DOI: 10.1116/1.581360

[27] Duc V. Dinh and Oliver Brandt 2024 Electrical properties of ScN(111) layers grown on semi-insulating GaN(0001) by plasma-assisted molecular beam epitaxy Phys. Rev. Applied 22, 014067 DOI: 10.1103/PhysRevApplied.22.014067

[28]R. P. Frankenthal *et al* 1983 J. Electrochem. Soc. 130 2056 DOI 10.1149/1.2119522

[29] W.J. Skocpol, M.R. Beasley, M. Tinkham, 1974 Self-heating hotspots in superconducting thin-film microbridges J. Appl. Phys. 45 (9) 4054 DOI: 10.1063/1.1663912

[30] Il'in, K., Siegel, M., Engel, A., Bartolf, H., Schilling, A., Semenov, A., & Huebers, H.-W. 2008 Current-Induced Critical State in NbN Thin-Film Structures. Journal of Low Temperature Physics, 151(1–2), 585–590. https://doi.org/10.1007/s10909-007-9690-5

[31] M. Yu. Kupriyanov , V. F. Lukichev 1980 Temperature dependence of pair-breaking current in superconductors J. Low Temp. Phys. 6, 210–214 1980 https://doi.org/10.1063/10.0030039

[32] A. Leitner et al Phys. Rev. B 62, 1408 (2000) DOI: 10.1103/PhysRevB.62.1408

[33] Vodolazov, D. Y. 2017 Single-Photon Detection by a Dirty Current-Carrying Superconducting Strip Based on the Kinetic-Equation Approach. Physical Review Applied, 7(3), 034014. https://doi.org/10.1103/PhysRevApplied.7.034014

[34] J.A.Thornton 1977 High rate thick film growth Ann. Rev. Mater. Sci. 7: 239-260 doi.org/10.1146/annurev.ms.07.080177.001323